\documentclass[twocolumn,showpacs,aps]{revtex4}

\usepackage{graphicx}
\usepackage{dcolumn}
\usepackage{bm}

\begin{document}
\title {Deviations from plastic barriers in Bi$_2$Sr$_2$CaCu$_2$O$_{8+\delta}$ thin films}
\author {Y. Z. Zhang}
\author{Z. Wang}
\author{X. F. Lu}
\author{H. H. Wen}
\affiliation{National Laboratory for superconductivity, Institute
of Physics \& Center for Condensed Matter Physics, Chinese Academy
of Sciences, P. O. Box 603,  100080, Beijing, China}
\author{J. F. de Marneffe}
\altaffiliation[Present address: ] {Interuniversitair
Microelectronica Centrum - IMEC v.z.w., 75 Kapeldreef, B-3001
Leuven, Belgium}
\author{R. Deltour}
\affiliation{Universit\'{e} Libre de Bruxelles, Physique des
Solides, CP-233, B-1050, Brussels, Belgium}
\author{A. G. M. Jansen}
 \author{P. Wyder}
\affiliation{Grenoble High Magnetic Field Laboratory,
Max-Planck-Institut f$\ddot{u}$r Festk$\ddot{o}$rperforschung and
Centre National de la Recherche Scientifique, 25, Avenue des
Martyrs, B. P. 166, F-38042 Grenoble Cedex 9, Grenoble, France}
\begin{abstract}
Resistive transitions of an epitaxial
Bi$_2$Sr$_2$CaCu$_2$O$_{8+\delta}$ thin film were measured in
various magnetic fields ($\textbf{H}\parallel \textbf{c}$),
ranging from 0 to 22.0 T. Rounded curvatures of  low resistivity
tails are observed in Arrhenius plot and considered to relate to
deviations from plastic barriers. In order to characterize these
deviations, an empirical barrier form is developed, which is found
to be in good agreement with experimental data and coincide with
the plastic barrier form in a limited magnetic field range. Using
the plastic barrier predictions and the empirical barrier form, we
successfully explain the observed deviations.

\end{abstract}

\maketitle

One of the most intriguing features of high $T_c$ superconductors
(HTSCs) is the remarkable broadening of resistive transitions in
applied magnetic fields. The broadening is related to thermal
barriers (thermal activation energies) for vortex motion. In
general, the vortex motion can be divided into three
characteristic regimes
\cite{Blatter,Brandt,Cohen,Geshkenbein,Vinokur}. In the high
temperature regime where the barrier  $U_0 \le T$, resistivity is
given by flux flow resistivity $\rho\propto B/H_{c2}$. In the
intermediate temperature regime, flux motion occurs through
thermally assisted flux flow (TAFF), where flux lines are weakly
pinned in the vortex liquid with $U_0\gg T$, and resistivity
$\rho\propto \exp(-U_0/T)$, where $U_0$ is independent of the
current density $j$ for $j\to 0$. In  the low temperature regime,
the form $\rho\propto \exp(-U_0/T)$ remains valid for the
resistivity analysis with $U_0(j)$ growing unlimitedly  for $j\to
0$, thus leading to $\rho\to 0$.

Bi$_2$Sr$_2$CaCu$_2$O$_{8+\delta}$ (Bi-2212) is a strongly anisotropic superconductor with a layered crystalline structure. The corresponding vortex \cite {Blatter,Brandt,Cohen} matter is highly  two-dimensional (2D) in high magnetic fields, and is three-dimensional (3D) in low magnetic fields. The study of the activation energy of Bi-2212 is very interesting, as its TAFF regime is very broad and gives the necessary knowledge for understanding the vortex characteristics in HTSCs. Generally, resistivity in the TAFF regime is often analyzed in an Arrhenius plot with the approximation $\ln \rho(T, H)\approx\ln\rho_0 -U_0/T$ \cite {Palstra}, where $\ln \rho_0$ is the logarithmic resistivity for linearly extrapolating to $1/T=0$, and $U_0$ is the average slope for the resistivity data in the low resistivity portion of the curves.  Palstra  \textit {et al.} \cite{Palstra} found a power law dependence $U_0\propto H^{-\alpha}$ with $\rho_0$ being several orders magnitude larger than the normal state resistivity in HTSCs. Kucera \textit {et al.} \cite {Kucera} suggested that the prefactor $\rho_0$ could be highly reduced with a factor $\exp(U_0/T_c)$ and that the activation
energy  $U_0\propto H^{-1/2}(1-T/T_c)$ for Bi-2212 thin films,
where $T_c$ was the critical temperature. The same relation
$U_0\propto H^{-1/2}(1-t)$ was also suggested by Wagner \textit
{et al.} \cite {Wagner} for Bi-2212 thin films, where $t=T/T_c$.

For explaining the vortex dynamics of HTSCs, many theoretical
approaches have been proposed to characterize the activation
energies \cite{Blatter, Brandt,Cohen, Yeshurun, Tinkham,
Feigelman, Palstra, Geshkenbein, Vinokur}.  Among these
approaches, the scaling of the barrier $U\propto H^{-1/2} (1-t)$
was first theoretically suggested by Geshkenbein et al. in 1989
\cite {Geshkenbein}, and then developed by Vinokur et al.
\cite{Vinokur}. This theory is  based on the model of plastic flux
creep ascribing the dissipation to the plastic shear of
dislocations in a weakly pinned vortex liquid. It seems that this
model perfectly describes the barrier relation of Bi-2212 thin
films determined by Kucera \textit {et al.} \cite {Kucera} and
Wagner \textit {et al.} \cite{Wagner}. However, this model is
based on the analysis of 3D vortex dynamics that provides a poor
correspondence with the highly 2D vortex matter for which vortex
cutting and reconnecting can change the plastic barriers in the
same order of magnitude \cite {Vinokur}.  Previously, most of the
published papers have extensive discussions on the regions of
validity of the plastic creep concept. Deviations of the concept
in experiments are observed \cite{Palstra, Kucera, Wagner, Zhang},
but have not  been studied detailedly until now. As a consequence,
a detailed study of the creep deviations from the plastic barrier
model predictions is of primary interest.

In this paper, we report measurements of resistive transitions of
a Bi-2212 thin film in magnetic fields parallel to $c$-axis from 0
to 22.0 T.  comparing these transitions with previously published
papers, we develop an empirical barrier form for describing the
deviations from plastic barriers.  We find that this empirical
form coincides with the plastic barrier form in a limited magnetic
field range. By using this new expression, we successfully explain
the observed deviations.

%
\begin{figure}[t]
\includegraphics
[width= .66\columnwidth] {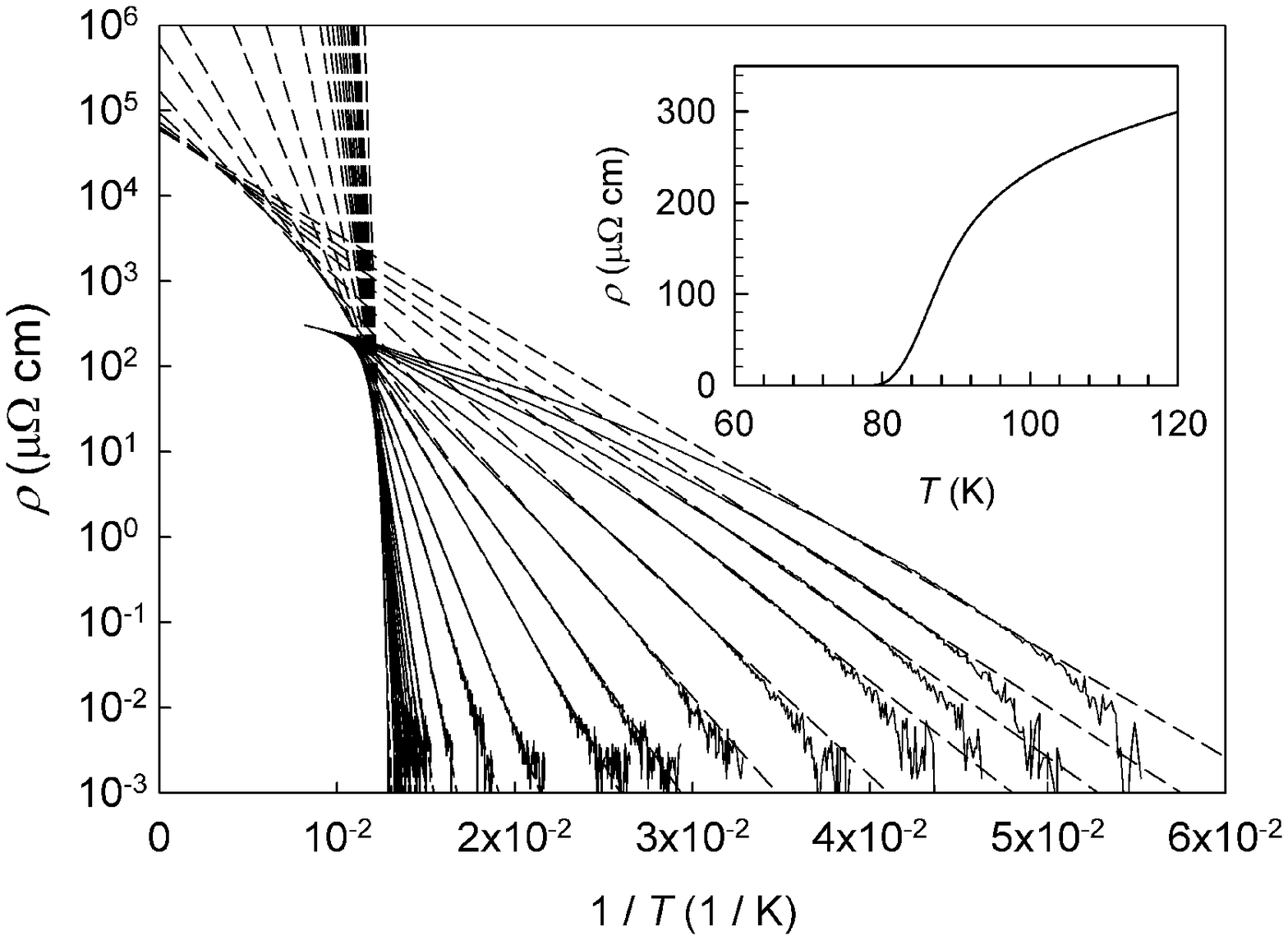} \caption{\label{f1}The Arrhenius
plot of the Bi$_2$Sr$_2$CaCu$_2$O$_{8+\delta}$ thin film. From
left to right: $\mu_0H=$ 0.0, 0.0037, 0.0052, 0.0070, 0.0089
0.013, 0.021, 0.030, 0.050, 0.078, 0.113, 0.157, 0.302, 0.604,
1.0, 2.0, 3.0, 5.0, 8.0, 12.0, 15.0, 18.0, 22.0 T. The dashed
lines are linear regressions of the data in the range
$10^{-4}\rho_n\le\rho\le 10^{-2}\rho_n$. The inset is the
$\rho(T,H=0)$ curve.}
\end{figure}

Epitaxial Bi-2212 thin films were prepared by an inverted cylinder
magnetron sputtering technique on (100) SrTiO$_3$  and (100)
LaAlO$_3$ substrates.  The composition of the target was
compensated in order to reach an ideal composition in the thin
films. The sputtering gas was a 1:1 mixture of Ar and O$_2$ at
$100$ Pa. Deposition temperature was in the range of $810 \sim
840^\circ$C. After deposition, Bi-2212 thin films were annealed in
an atmosphere of 10 Pa pure O$_2$ at $\sim 500^\circ$C for 45 min.
X-ray diffraction patterns show that thin films are highly
$c$-axis oriented and epitaxial. The studied film with a thickness
of  210 $\pm 20$ nm was patterned with a microbridge [500  $\mu$m
(length) $\times$ 100 $\mu$m (width)].  Gold leads were stuck onto
the film with silver paste. In order to reduce the resistance
between the film and the gold wires, the film was baked at
350$^\circ$C in flowing oxygen for 6 hrs.  Bipolar DC current of
40 $\mu$A (corresponding to the current density of $\sim 190$
A/cm$^2$)  was applied for the resistive measurement. This current
density ensures that the low resistivity  is ohmic in the most
range for the measurement \cite {Wagner}.

Figure~\ref{f1} shows $\rho(T,H)$ data of the Bi-2212 thin film in
an Arrhenius plot.  The dashed lines in Fig.~\ref{f1} are linear
regressions for the resistivity data range of
$10^{-4}\rho_n\le\rho\le 10^{-2}\rho_n$, where $\rho_n=\rho(120$
K$)\approx 300$ $\mu \Omega$cm. A detailed examination of each
curve suggests that these regressions are in good agreement with
three or four order of magnitude of the resistivity data in a
limited magnetic field range ($0.021\le\mu_0 H\le 1.0$ T), but do
only approximately average the rounded curvatures for the other
ranges. In following discussion, we will simply use some special
field values as just mentioned  above, which are arbitrarily
defined by the intended field values in  measurements, as these
values shall be  close to the precise characteristic field values
of the sample in reality and give very close information about the
vortex matter.

\begin{figure}
\includegraphics
[width= .66\columnwidth] {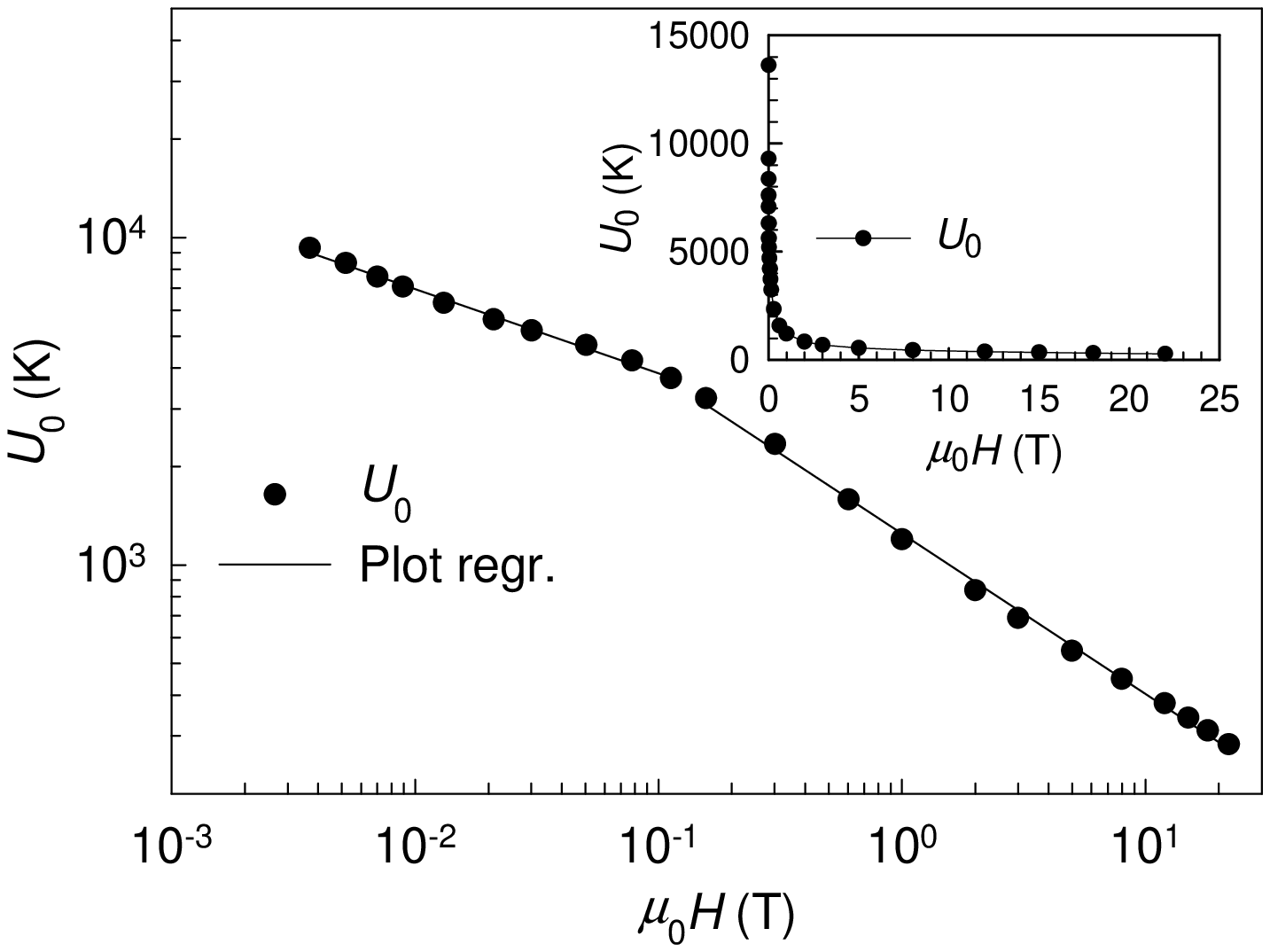} \caption{\label{f2} Magnetic field
dependence of $U_0$ in both of the double $\log$ scale and the
double linear scale (inset). The solid lines are plot regressions
of $U_0$ for two different field regimes.}
\end{figure}

Figure~\ref{f2} shows $U_0(H)$ data. The linear regressions of
$U_0(H)$ in the plot suggest a power law dependence $U_0\propto
H^{-\alpha}$ with $\alpha\approx 0.258$ for $\mu_0H\le 0.113$ T,
and $\alpha\approx 0.490$ for $\mu_0H\ge 0.157$ T. The second
$\alpha$ value consists with the plastic barrier form and the
results determined in Refs.~\cite {Kucera,Wagner}.  However,
rounded curvatures in the low resistivity portions are observed in
the Arrhenius plot for $\mu_0 H<0.021$ T and $\mu_0 H>1.0$ T,
which are apparently not described by the plastic barrier form.

\begin{figure}
\includegraphics
[width= .66\columnwidth] {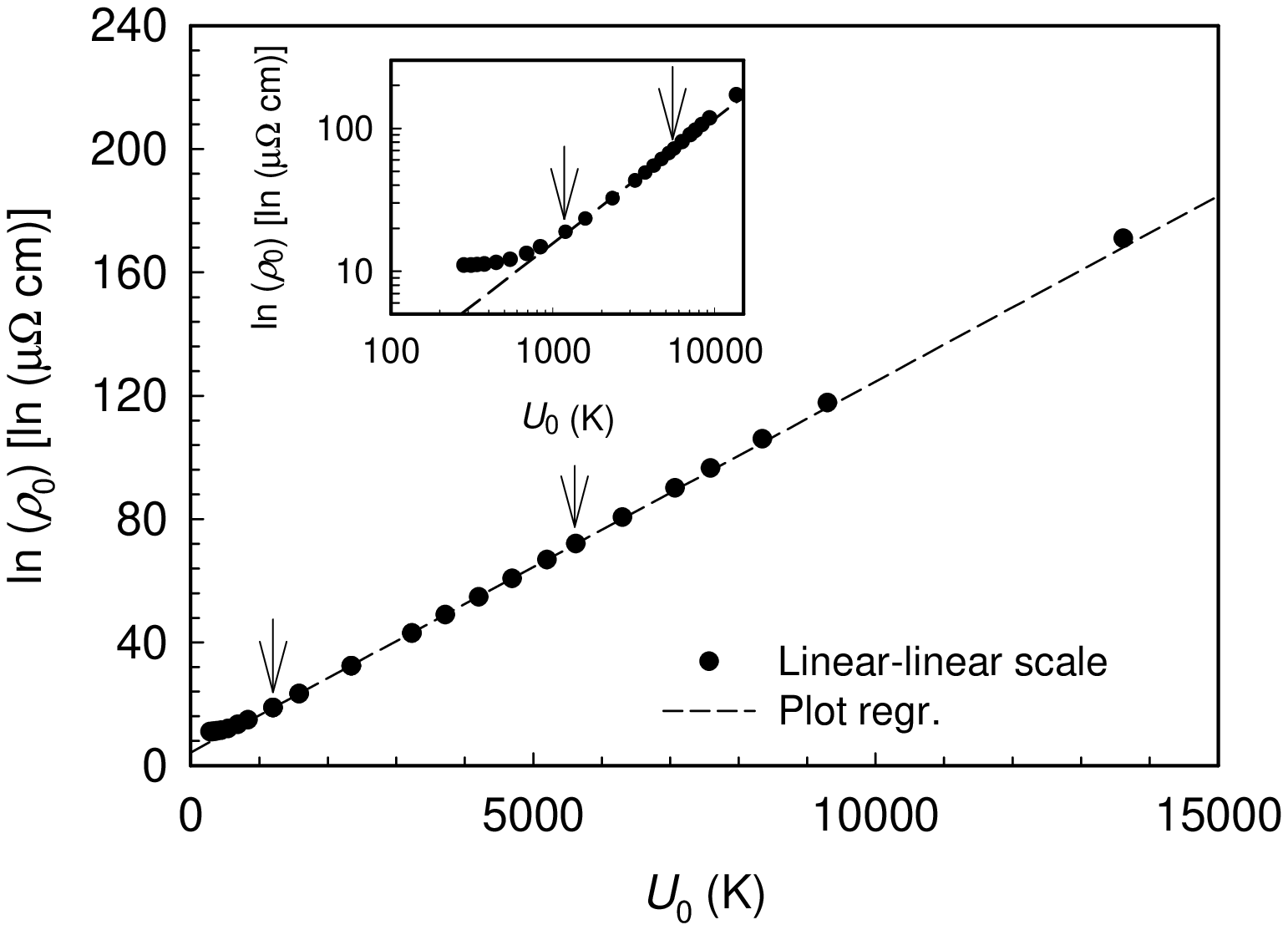} \caption{\label{f3} $\ln
\rho_0(U_0)$ data in both of the linear-linear scale  and the
$\log$-$\log$ scale (inset).  The dashed lines represent the plot
linear regressions for $1190<U_0<5620$ K (as denoted by the
arrows). }
\end{figure}

Figure~\ref{f3} shows the $\ln \rho_0(U_0)$ relation in both
linear-linear and log-log scales. Note that $\ln \rho_0(U_0)$ is
approximately linear for $1190<U_0<5620$ K corresponding to the
field  regime of $0.021\le\mu_0 H\le 1.0$ T, where the regressions
in the Arrhenius plots as shown in Fig.~\ref{f1} are also linear,
so that the determinations of $\ln \rho_0(U_0)$ are quite accurate
through the regime, and can be assuredly used to deduce some
important information as discussed below.

Considering the fact that many authors suggested $U_0\propto
(1-t)^\beta$ with $\beta=1$ in Refs.~\cite{Geshkenbein, Palstra,
Vinokur, Kucera, Wagner, Andersson, Thopart, Figueras},
$\beta=1.5$ in Refs.~\cite {Palstra, Yeshurun, Tinkham}, $\beta=2$
in Refs.~\cite{Yeshurun, Palstra}, and the $\beta$ value selected
from 1.5 to 2.4 in Ref.~\cite {Kim}, we start by assuming that
$\rho=\rho_{0f}\exp[-U(T, H)/T]$, where $\rho_{0f}$ is constant,
$U(T,H)=g(H)f(t)$, $g$ is the magnetic field dependence,
$f=(1-t)^\beta$, and $\beta$ accounts for the nonlinearity in the
Arrhenius plot. Using the progression $(1-t)^\beta= 1-\beta
t+\beta(\beta-1)t^2/2!-\beta(\beta-1)(\beta-2)t^3/3!+\ldots$, we
obtain $\ln \rho \approx (\ln \rho_{0f}+
g\beta/T_c)-(g/T)[1+\beta(\beta-1)t^2/2!-\beta(\beta-1)(\beta-2)t^3/3!+\ldots]$,
where the term $(\ln \rho_{0f}+ g\beta/T_c)\approx \ln \rho_0$ is
temperature independent. With $\beta= 1$, we have $\ln \rho_0
\approx \ln \rho_{0f}+ U_0/T_c$ as observed in the linear part of
Fig.~\ref{f3} for $1190<U_0<5620$ K (denoted by arrows), where
$U_0=g$. Here, the linear  $\ln\rho(U_0)$ portion corresponds to
the field range of  $0.021 \le \mu_0H\le 1.0$ T. By linearly
extrapolating $\ln \rho_0(U_0)$ to $U_0=0$, we find that
$\rho_{0f}\approx 69.7$ $\mu\Omega$cm, and $T_c$ $\approx 83.1$ K
is the approximation of the inverse value of the slope in the
double linear scale. Assuming $\beta=const$, a linear $\ln
\rho_0(U_0)$ relation will be found. Obviously, $\beta=const$
(including $\beta=1$) can not account for the nonlinear portions
of  $\ln \rho_0(U_0)$ curves in the regimes of $U_0<1190$ K and
$U_0>5620$ K.  It is interesting to note that if $\beta$ is
magnetic field dependent, $f$ becomes magnetic field dependent,
and thus a non-linear character is introduced into the $\ln
\rho_0(U_0)$ dependence.

\begin{figure}
\includegraphics
[width= .66\columnwidth] {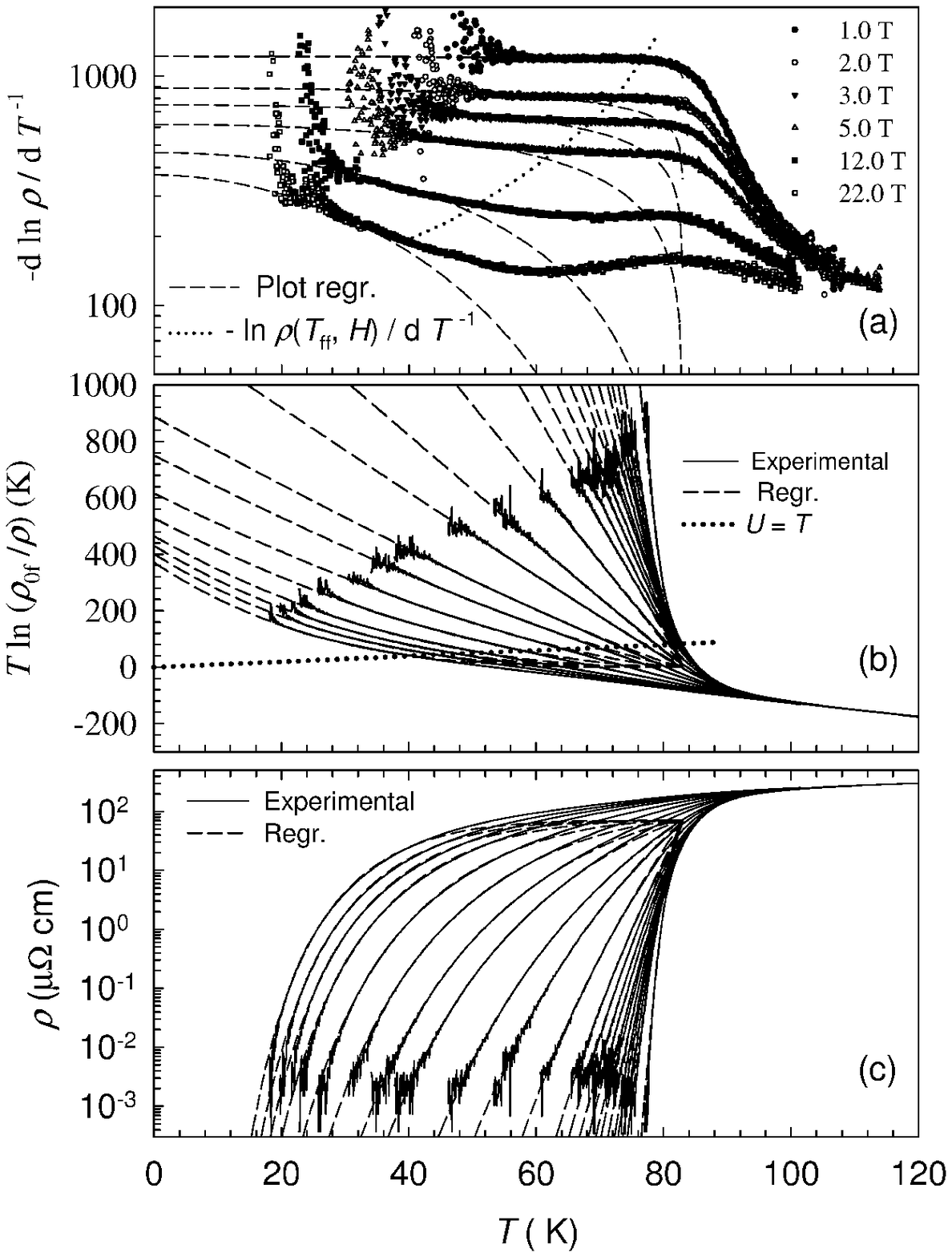} \caption{\label{f4} (a) The
different symbols give $-\partial \ln \rho/\partial T^{-1}$ data
in several magnetic fields as denoted by corresponding symbols.
The dotted line is the flux flow boundary determined in (b). (b)
The solid lines present $U(T,H)\approx T\ln [\rho_{0f}/\rho(T,H)]$
data for all the tested magnetic fields. The dotted line is $U=T$
corresponding to the flux flow boundary. (c) The solid lines are
$\rho(T,H)$ data for all the fields. The dashed lines in (a), (b),
and (c) are regressions using the empirical barrier form with the
same $g(H)$ and $\beta(H)$ (see text).}
\end{figure}

Previously, a magnetic field dependent $f$ was proposed by Palstra
\textit {et al.} \cite{Palstra} and  Kim \textit{et al.} \cite
{Kim} by introducing a magnetic field dependent $T_x(H)$ instead
of $T_c$ for the barrier scaling. Assuming $T_x = T_c/\beta$, we
find $\ln \rho_0 \approx \ln\rho_{0f}+g/T_x$ for the similar
explanation of the nonlinear $\ln \rho_0(U_0)$. In Bi-2212 thin
films, Kucera \textit {et al.} \cite{Kucera} and Wagner \textit{et
al.} \cite{Wagner} suggested that the barriers should \textbf
scale according to $U\propto H^{-1/2}(1-t)$ with a constant $T_c$
in $t$. However, $U(T,H)$ data for the low resistivity portion as
mentioned by the authors in Fig.~4 of Ref.~\cite {Kucera} do more
favor $T_x(H)$ than $T_c$. Fig.~\ref{f4}(a) shows $-\partial \ln
\rho(T,H)/\partial T^{-1}$ data with different symbols for
different magnetic fields. In the field range $0.021 \le\mu_0H\le
1.0$ T (not all shown in the figure for clarity), the data are
roughly temperature independent in the TAFF regime, indicating
that $U\propto (1-t)$. Note that $f=1-t$ will lead to $-\partial
\ln \rho/\partial T^{-1} = U-T\partial U/\partial T = g$,  where
the $g$ is temperature independent. For $\mu_0H>1$ T in the TAFF
regime, $-\partial \ln \rho/\partial T^{-1}$ of our Bi-2212 thin
film in Fig.~\ref{f4}(a) and Bi-2212 crystals in
Ref.~\cite{Palstra} are temperature dependent. It seems that
similar temperature dependences can also be deduced from high
field data in Refs. \cite {Kucera,Wagner}. These temperature
dependences do not support the $f=1-t$ argument even by
substituting $T_x(H)$ for $T_c$.

\begin{figure}
\includegraphics
[width= .66\columnwidth] {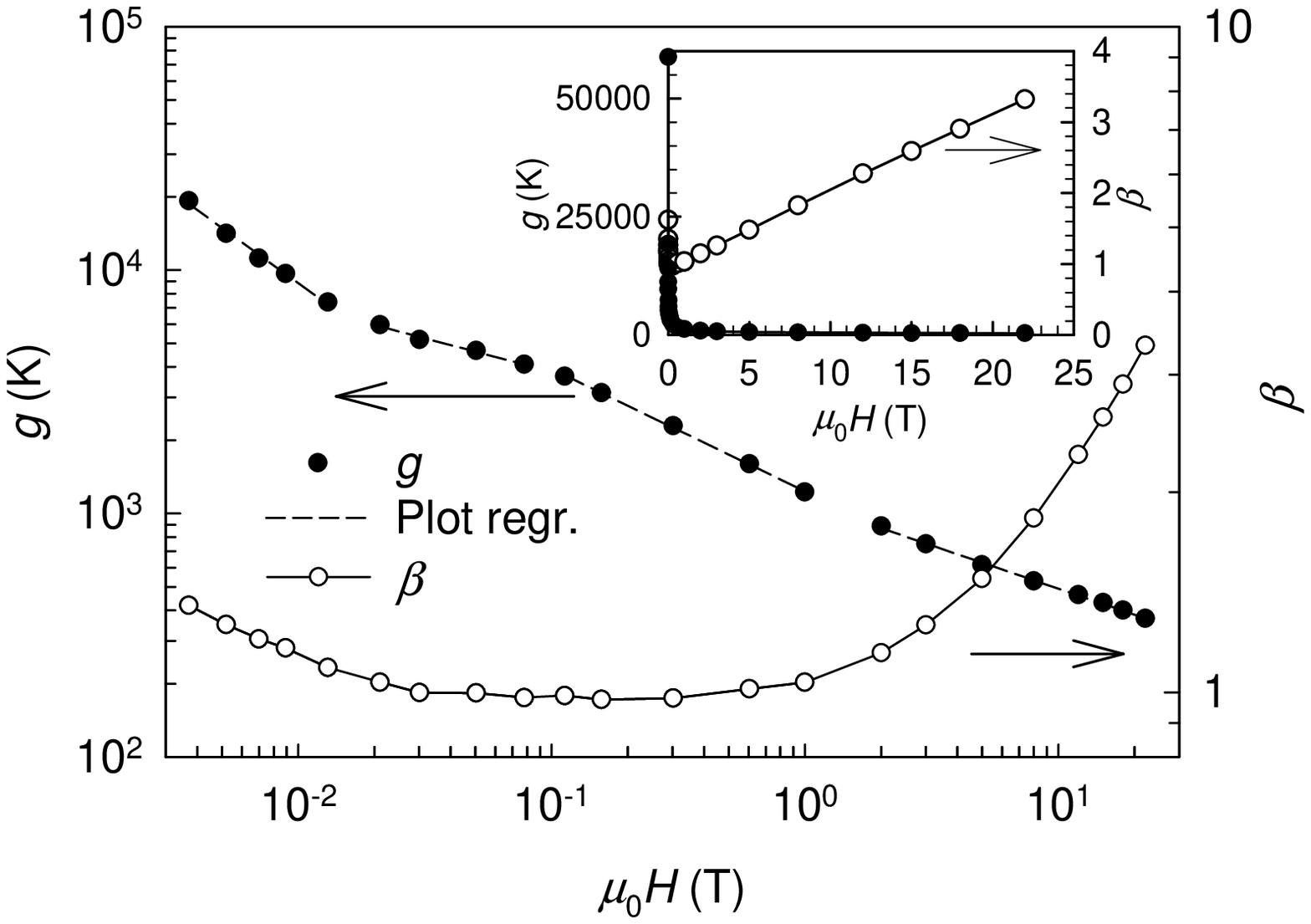}\caption{\label{f5} Solid circles
present $g(H)$ data and open circles show $\beta(H)$ data. The
dashed lines correspond to the plot regressions in the
$\log$-$\log$ scale. The inset shows $g(H)$ and $\beta(H)$ data in
linear-linear scales.}
\end{figure}

For high fields, each $-\partial \ln \rho/\partial T^{-1}$
monotonously decreases with temperature from low temperature
to a local minimum. One may take the data around these minima for
$U$ simulation in which a constant $\beta$ for $-\partial \ln
\rho(T,H)/\partial T^{-1}$ data may be determined \cite
{Kim1,Yang}.  However, taking into account the flux flow condition
$U\le T$, this may lead to a wrong result. Fig.~\ref{f4}(b) shows
$U(T, H)= T\ln [\rho_{0f}/\rho(T,H)]$ data with solid lines and
the flux flow boundary $U=T$ with the dotted line. The flux flow
temperature $T_{ff}(H)$ can  be determined with the crossing
points between  the $U(T,H)$  lines and the dotted line $U=T$.  We
thus draw the flux flow boundary,  $-\partial
\ln\rho(T_{ff},H)/\partial T^{-1}$, with a dotted line in
Fig.~\ref{f4}(a). It is found that minima of $-\partial \ln
\rho(T,H)/\partial T^{-1}$ may result in a corresponding
temperature higher than $T_{ff}$ for high fields. This means that
the determination of  a constant $\beta$ around the minima of
$-\partial \ln \rho(T,H)/\partial T^{-1}$ shall be dismissed.

As a result, we argue that we have to use  $f=(1-t)^\beta$ as a
substitute for $f=(1-t)$ in the barrier definition, where $\beta$
is magnetic field dependent. The dashed lines in Fig.~\ref{f4}(b)
correspond to the best regressions using the expression $U(T,
H)=g(1-t)^\beta$ for which the resistivity data in the range of
$10^{-4}\rho_n\le\rho\le 10^{-2}\rho_n$ are used,  where $g$ and
$\beta$ are free fitting parameters.  We also present dashed lines
using the same $g(H)$ and $\beta(H)$ for $-\partial \ln
\rho(T,H)/\partial T^{-1}$ and $\rho(T,H)$ in Fig.~\ref{f4}(a) and
(c), respectively. These regressions are  in good agreement with
$U(T,H)$,  $-\partial \ln \rho(T,H)/\partial T^{-1}$, and
$\rho(T,H)$ in the TAFF regime.

Figure~\ref{f5} shows $g(H)$ and $\beta(H)$ data, respectively.
From the figure, we can roughly divide the $g$ data into four
magnetic field regimes according to the field values that were
used in the measurements. We find that both $g(H)$ and $\beta(H)$
have an apparent increase for $\mu_0H\le 0.013$ T where
$\alpha\approx 0.751$, and $\beta$ increases with decreasing
field, indicating a deviation from the plastic barrier model. As
mentioned in many articles \cite {Kosterlitz,Halperin,Minnhagen},
the binding and unbinding behaviors of 2D vortex-antivortex pairs
dominate the low resistivity in the low magnetic field range.
Obviously, the 2D behaviors do not relate to the plastic vortex
motion.  In the range of $0.021\le \mu_0H\le 1.0$ T, $\beta\approx
1$,  $\alpha \approx 0.275$ for $0.021\le \mu_0H \le 0.113$ T, and
$\alpha \approx 0.502$ for $0.157\le \mu_0H\le 1.0$ T.  For
$0.021\le\mu_0H\le 0.113$ T, the intervortex spacing is relatively
large and the vortex matter is in a 3D state  where the vortex
system is very close to or can be in the plastic barrier regime
\cite {Kucera, Wagner}.  For $0.157\le \mu_0H \le 1.0$ T, both
$\alpha$ and $\beta$ have the values predicted by the plastic
form, indicating that the vortex system is in the plastic barrier
regime. Note that the vortex system changes from 3D to 2D at a
crossover field $\mu_0H_d\approx 4\phi_0/\gamma^2d^2$, where
$\gamma$ is the anisotropic factor with $50\le\gamma\le 200$ in
Bi-2212 \cite{Blatter, Brandt, Cohen}, $d$ is the interplanar
spacing, and $\phi_0$ is the flux quantum. If
$0.157\le\mu_0H<\mu_0H_d$,  the vortex system is 3D for the
plastic barriers. If $\mu_0H_d<\mu_0H<1$ T,  the system is in a 2D
state where it maintains some 3D characteristics allowing plastic
barrier behaviors. These 3D characteristics are gradually
destroyed by further increasing the magnetic field ($\mu_0H>1.0$
T), where $\alpha\approx 0.355$ and $\beta$ increases with
$\mu_0H$ as shown in Fig.~\ref{f5}. For $\mu_0H>1.0$ T, the vortex
matter gradually crosses over into a highly 2D state where 2D
vortices (pancake vortices) are largely overlapped and 2D
collective interaction dominates the vortex behaviors; besides,
the plastic vortex behavior has to fade away due to a strong
interlayer decoupling \cite {Blatter, Brandt, Cohen, Geshkenbein,
Vinokur,Zhang}. In particular, Kucera \textit{et al.} \cite
{Kucera} and Wagner \textit{et al.} \cite{Wagner} also mentioned
deviations of the plastic barriers at high magnetic fields which
were suggested to relate to a 3D to 2D transition.

Note that both $U(T,H)$ and $-\partial \ln \rho(T,H)/\partial
T^{-1}$ increase with decreasing temperature and deviate from the
regressions in  low temperature. This implies that the vortex
coupling and pinning are enhanced.  The deviations corresponding
to the curvature differences and the curve separations between
experimental data and fittings are a consequence of changes of
competitive relations between pinning and depinning, and between
coupling (reconnecting) and decoupling (cutting). These changes
may gradually drive $U(T,H)$ into the $j$ dependent regime with
decreasing temperature for $j\to 0$.

It is easily found that the barrier estimations with the empirical
and the plastic barrier forms ($g$ in Fig.~\ref{f5} and $U_0$ in
Fig.~\ref{f2}) have the same order that is just consistent with
the plastic barrier prediction for any vortex deformation \cite
{Vinokur}. In this case, the empirical form coincides with the
plastic barrier prediction.  The similar barrier relation and
values, obtained by AC susceptibility measurements of a similar
Bi-2212 thin film for $\mu_0H\le 1.0$ T, give  a support to the
$g(H)$ determination \cite {Marneffe}.

Note that the increasing $\beta$ ($\beta>1$) is a common behavior
with increasing 2D feature for $\mu_0H\le 0.013$ T  and
$\mu_0H>1.0$ T.  This  implies that the increasing $\beta$
features, as shown in Fig.~\ref{f5}, give the signs of a crossover
from 3D to 2D, which differs on both field sides by its strength.
For $H\to 0$, the low resistivity portion is dominated by the 2D
behaviors of binding and unbinding of vortex-antivortex pairs. In
high field,  influences of interlayer decoupling and 2D collective
behaviors must be taken into account for increasing $H$.

In summary, based on experimental results, we have developed an
empirical barrier form $U\propto H^{-\alpha(H)}(1-t)^{\beta(H)}$
in Bi-2212 thin films.  This expression coincide with the plastic
barrier prediction over the magnetic field range $0.021\le \mu_0H
\le 1.0$ T, and can be applied to account for the deviations from
plastic barriers in Bi-2212 thin films. Moreover, this model may
possibly be used for the analysis of TAFF behaviors in other
HTSCs.

This work has been financially supported by the National Science
Foundation of China (Serial No. 10174091 and 10174093).

\end{document}